\documentclass[twocolumn]{jpsj2}
\usepackage{graphicx}

\def\runtitle{
Quantum Narrowing Effect in a Spin-Peierls System
with Quantum Lattice Fluctuation
}
\def\runauthor{
Hiroaki \textsc{Onishi} and Seiji \textsc{Miyashita}
}

\title{
Quantum Narrowing Effect in a Spin-Peierls System\\
with Quantum Lattice Fluctuation
}

\author{
Hiroaki \textsc{Onishi}$^{1,2,3}$ and Seiji \textsc{Miyashita}$^1$
}

\inst{
$^1$%
Department of Applied Physics,
Graduate School of Engineering,
University of Tokyo,\\
7-3-1 Hongo, Bunkyo-ku, Tokyo 113-8656\\
$^2$%
Advanced Science Research Center,
Japan Atomic Energy Research Institute,\\
2-4 Shirakata Shirane, Tokai-mura, Naka-gun, Ibaraki 319-1195\\
$^3$%
Department of Earth and Space Science,
Graduate School of Science,
Osaka University,\\
1-1 Machikaneyama-cho, Toyonaka, Osaka 560-0043
}

\recdate{
Apr 3, 2003
}

\abst{
We investigate
a one-dimensional $S=1/2$ antiferromagnetic Heisenberg model
coupled to quantum lattice vibration
using a quantum Monte Carlo method.
We study the ground-state lattice fluctuation
where the system shows a characteristic structure factor.
We also study the mass dependence of magnetic properties
such as the magnetic susceptibility and the magnetic excitation spectrum.
For heavy mass,
the system shows the same behavior as
the case of classical lattice vibration.
On the other hand,
for light mass,
magnetic properties coincide with those of
the static uniform chain.
We investigate the physical mechanism of this behavior
and propose the picture of quantum narrowing.
}

\kword{
spin-Peierls transition,
quantum lattice fluctuation,
quantum narrowing effect,
quantum Monte Carlo method,
loop algorithm
}

\begin{document}
\sloppy
\maketitle

\section{Introduction}
\label{sec:introduction}

The spin-Peierls transition is one of the most fascinating features
derived from
cooperative phenomena of spin and lattice degrees of freedom
due to the spin-phonon coupling
in the one-dimensional antiferromagnetic
Heisenberg model.~\cite{review-ELCC}
The lattice shows a spontaneous dimerization
which accompanies the bond alternation
at low temperature,
because
a pair of spins tends to form a local singlet state on each strong bond
to lower the magnetic energy
which competes with the elastic energy of the lattice.
An energy gap opens
in the spin excitation spectrum.
The magnetic susceptibility in all directions
exponentially drops to zero
due to the spin gap
as the temperature decreases.
These phenomena have been observed
in materials
such as TTFCuS$_4$C$_4$(CF$_3$)$_4$~\cite{TTFCuBDT}
and CuGeO$_3$.~\cite{CuGeO3}

Theoretical investigations have revealed
ground state properties
with an adiabatic approximation for the lattice.
Usually,
the exchange coupling is assumed to change
in proportion to the lattice distortion amplitude $\epsilon$.
Namely,
the bond alternation is given in the form
$J_i=J[1+(-1)^i\delta]$ and $\delta=\alpha\epsilon$.
It has been shown
by a bosonization method~\cite{Cross-Fisher,Nakano-Fukuyama}
that
the bond alternation brings about the magnetic energy gain
proportional to $\delta^{4/3}$ for $S=1/2$,
which overcompensates the elastic energy loss proportional to $\delta^2$.
Thus
the system dimerizes in the ground state
for an arbitrarily small spin-phonon coupling.
The adiabatic treatment for the lattice is valid
when the mass of the magnetic ion is so heavy
that the kinetic energy of the lattice vibration can be neglected.
Put another way,
the phonon energy $\omega$ is much small compared to
the spin-Peierls gap and the exchange coupling.
Recently we have studied
finite temperature properties
of the adiabatic lattice case
neglecting the kinetic energy term of the lattice
by a quantum Monte Carlo method.~\cite{spqmc}
We have observed how the bond alternation develops
as the temperature decreases.

On the other hand,
quantum lattice fluctuation becomes remarkable for the light mass case.
It has been revealed that
quantum lattice fluctuation
disarranges the lattice dimerization in the ground state
below a critical spin-phonon coupling.~\cite{ph-dmrg1,ph-a1,ph-a2}
The quantum phase transition
between the dimerized phase and the uniform phase has been investigated
with various methods
such as an exact diagonalization method,~\cite{ph-ed}
a density matrix renormalization group method,~\cite{ph-dmrg1,ph-dmrg2}
a quantum Monte Carlo method~\cite{ph-qmc1,ph-qmc2,ph-qmc3}
and an analytical method.~\cite{ph-a1,ph-a2,ph-a3,ph-a4}
In analytical investigations,
the lattice degree of freedom is integrated out for small $\omega$
and the effect of the spin-phonon coupling is transposed to
the dimerization and the geometrical frustration of the spin interaction
in an effective spin Hamiltonian.~\cite{ph-a2,ph-a3}
Thermodynamic properties have been investigated
by a quantum Monte Carlo method
taking account of thermal fluctuation
of the quantum phonon.~\cite{ph-qmc1,ph-qmc3}
For small $\omega$,
the lattice fluctuation can be regarded as the adiabatic motion
and the lattice dimerization occurs at low temperature.
There the magnetic susceptibility decays exponentially.
In addition,
it has been pointed out that
thermal fluctuation of the lattice causes
deviation of the magnetic susceptibility from that of the static uniform chain
even at high temperature
where the spin-Peierls correlation is not relevant.
As $\omega$ becomes larger,
the magnetic susceptibility approaches
that of the original spin chain of the static lattice,
which suggests that
spin and lattice degrees of freedom decouple
and lattice fluctuation does not affect the spin state.

In this paper,
we pay attention to
the mass dependence of the effect of quantum lattice fluctuation.
In \S\ref{sec:model and method}
we explain a model and a numerical method.
In \S\ref{sec:lattice configuration}
we study the mass dependence of the lattice configuration.
For heavy mass,
the lattice is uniform on the thermal average at high temperature
and dimerizes at low temperature.
For light mass,
the spin-Peierls correlation does not appear even at low temperature.
We find distinctive dependence of the structure factor on the mass.
In \S\ref{sec:quantum lattice fluctuation}
we study the effect of quantum lattice fluctuation
from a viewpoint of the world-line configuration of the lattice.
In \S\ref{sec:mass dependence of magnetic properties}
we study the mass dependence of magnetic properties
such as the magnetic susceptibility
and the magnetic excitation spectrum.
For light mass,
we find no effect of lattice fluctuation on magnetic properties,
even though the lattice strongly fluctuates.
In \S\ref{sec:quantum narrowing effect}
we study the reason why
the largely fluctuating lattice
gives the same magnetic behavior as that of the static uniform chain.
It can be interpreted
as a narrowing effect due to quantum lattice fluctuation.
In \S\ref{summary and discussion}
we summarize and discuss our results.

\section{Model and Method}
\label{sec:model and method}

We investigate
a one-dimensional $S=1/2$ antiferromagnetic Heisenberg model
coupling with quantum lattice vibration,
in which the lattice displacement produces
the distortion of the exchange coupling between spins.
The Hamiltonian is described by
\begin{eqnarray}
&&
H
=
\sum_{i=1}^{N}
J\left[1+\alpha(u_{i}-u_{i+1})\right]
\mbox{\boldmath $S$}_i\cdot\mbox{\boldmath $S$}_{i+1}
\nonumber\\
&&
\ \ \ \ \ \ 
+
\sum_{i=1}^{N}
\left[
\frac{1}{2m}p_{i}^2 + \frac{k}{2}(u_{i}-u_{i+1})^2
\right],
\label{eq:hamsp}
\end{eqnarray}
where
the distortion of the exchange coupling is assumed to be proportional to
the lattice distortion.
Here
$\alpha$ is the spin-phonon coupling constant,
$m$ is the mass of the magnetic ion
and $k$ is the elastic constant.
The periodic boundary condition is adopted
for both spin and lattice degrees of freedom.
In the simulation,
we fix the total length of the lattice.
We set the uniform exchange coupling $J=1$ and take it as the energy unit.
We consider a parameter set of $\alpha=1$ and $k=1$
and investigate the behavior for various values of $m$.

%
%
\begin{figure}[b]
\begin{center}
\includegraphics[width=60mm]{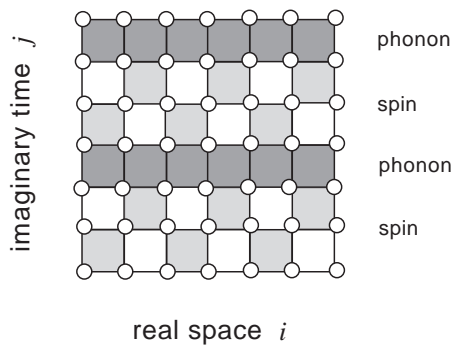}
\end{center}
\caption{%
The schematic representation
of the ($1+1$)-dimensional lattice of $N=6$ and $M=2$.
}
\label{fig:cbdec}
\end{figure}

We use a quantum Monte Carlo method
in order to investigate thermodynamic properties.
We adopt the recipe introduced by Hirsch
to deal with the spin-phonon coupled system.~\cite{Hirsch-1,Hirsch-2}
The partition function is expressed in the path-integral formula
including both spin and lattice degrees of freedom,
\begin{eqnarray}
&&
\hspace*{-8mm}
Z
=
{\rm Tr}_{\rm spin} \left(\prod_{i,j}\int{\rm d}u_{i,j}\right)
\prod_{j=1}^{M}
\nonumber\\
&&
\hspace*{-6mm}
\times
\exp\left[
-\Delta\tau
\sum_{i={\rm odd}}
J\left[1+\alpha(u_{i,j}-u_{i+1,j})\right]
\mbox{\boldmath $S$}_{i,j}\cdot\mbox{\boldmath $S$}_{i+1,j}
\right]
\nonumber\\
&&
\hspace*{-6mm}
\times
\exp\left[
-\Delta\tau
\sum_{i={\rm even}}
J\left[1+\alpha(u_{i,j}-u_{i+1,j})\right]
\mbox{\boldmath $S$}_{i,j}\cdot\mbox{\boldmath $S$}_{i+1,j}
\right]
\nonumber\\
&&
\hspace*{-6mm}
\times
\exp\left[
-\Delta\tau
\sum_{i=1}^{N}
\left[
\frac{m}{2}\left(\frac{u_{i,j+1}-u_{i,j}}{\Delta\tau}\right)^2
\right.\right.
\nonumber\\
&&
\hspace*{20mm}
\left.\left.
+ \frac{k}{2}(u_{i,j}-u_{i+1,j})^2
\right]
\right],
\end{eqnarray}
where $M$ is the Trotter number and $\Delta\tau=\beta/M$
with the inverse temperature $\beta$.
In Fig.~\ref{fig:cbdec}
we show an example of the transformed ($1+1$)-dimensional lattice.
There are three layers in a unit for the imaginary-time axis,
namely,
two checkerboard layers for the spin-spin part
and one layer for the phonon part.
The unit of layers is heaped in $M$ layers.

%
%
\begin{figure}[t]
\begin{center}
\includegraphics[width=60mm]{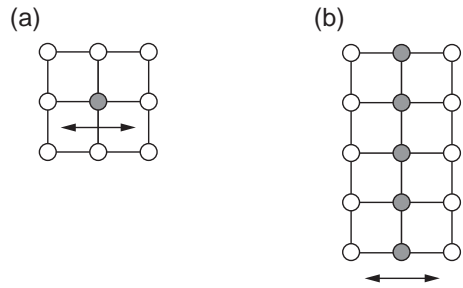}
\end{center}
\caption{%
The schematic representation of the change of the lattice configuration.
(a) the local flip and (b) the global flip
in which the lattice points denoted by solid circles are flipped together.
}
\label{fig:phflip}
\end{figure}

We perform Monte Carlo simulations
in the ($1+1$)-dimensional classical system.
The spin configuration is updated by the loop algorithm~\cite{loop-1,loop-2}
with the fixed lattice configuration.
The loop algorithm is powerful to calculate physical quantities
at considerable low temperature.
Therefore it is useful to observe the spin-Peierls correlation
at low temperature.
In the loop algorithm,
a graph element is assigned to each plaquette with a probability
which depends on the value of the exchange coupling on the plaquette.
In the present case,
the value of the exchange coupling varies along the imaginary-time axis
due to quantum lattice fluctuation.
Hence
we can not take the limit of continuous imaginary time.~\cite{loop-c}
In this study,
the lattice configuration is updated by the Metropolis algorithm
with the fixed spin configuration.
We perform a local flip and a global flip for the update of the lattice
as shown in Fig.~\ref{fig:phflip}.
The local flip updates the lattice state
at a point in the ($1+1$)-dimensional space,
which causes a curved world line of the lattice configuration.
For convenience of the calculation,
the change of the lattice displacement is discretized
by a small value $u_{\rm unit}$.
First,
whether the lattice point moves right or left is chosen
with a probability $1/2$.
Next,
whether the move takes place or not is chosen
according to the Boltzmann weight.
The local flip is efficient for the case where
quantum lattice fluctuation is large.
When quantum lattice fluctuation is small
and the lattice tends to move adiabatically,
we adopt the global flip of the lattice state.
The global flip updates the lattice states
by a parallel shift of all the points in a world line.
We perform the spin update and the lattice update alternately.
Therefore we obtain the thermal distribution
of both spin and lattice degrees of freedom.
Typically,
the initial $10^5$ Monte Carlo steps (MCS) are discarded
for thermalization,
and the following $10^6$ MCS are used
to calculate physical quantities.
The sampled data is divided into 10 bins,
and the errorbar is estimated from the standard deviation
for the data set of these 10 bins.

For the simplicity of the notation,
we use the bond distortion,
\begin{equation}
\Delta_{i}=\alpha(u_{i}-u_{i+1}),
\end{equation}
as a variable
instead of the lattice displacement $u_i$ itself.
We provide a cutoff for the bond distortion
$|\Delta_i|\leqslant\Delta_{\rm cutoff}=0.6$,
i.e., $0.4\leqslant J_i \leqslant 1.6$,
to confine the exchange coupling to be antiferromagnetic.
As we have pointed out,~\cite{spqmc}
the variation of the bond is very large at finite temperature
if we use the Hamiltonian of eq.~(\ref{eq:hamsp}).
Such large deviation is not realistic
and we need some nonlinear suppression of the bond deviation.
The cutoff plays the role of this suppression.

\section{Lattice Configuration}
\label{sec:lattice configuration}

The bond configuration characterizes the spin-Peierls state,
because
the bond takes
either a uniform configuration in the uniform state
or an alternating configuration in the spin-Peierls state.
We investigate the effect of both thermal fluctuation and quantum fluctuation
on the bond configuration.

%
%
\begin{figure}[t]
\begin{center}
\includegraphics[width=80mm]{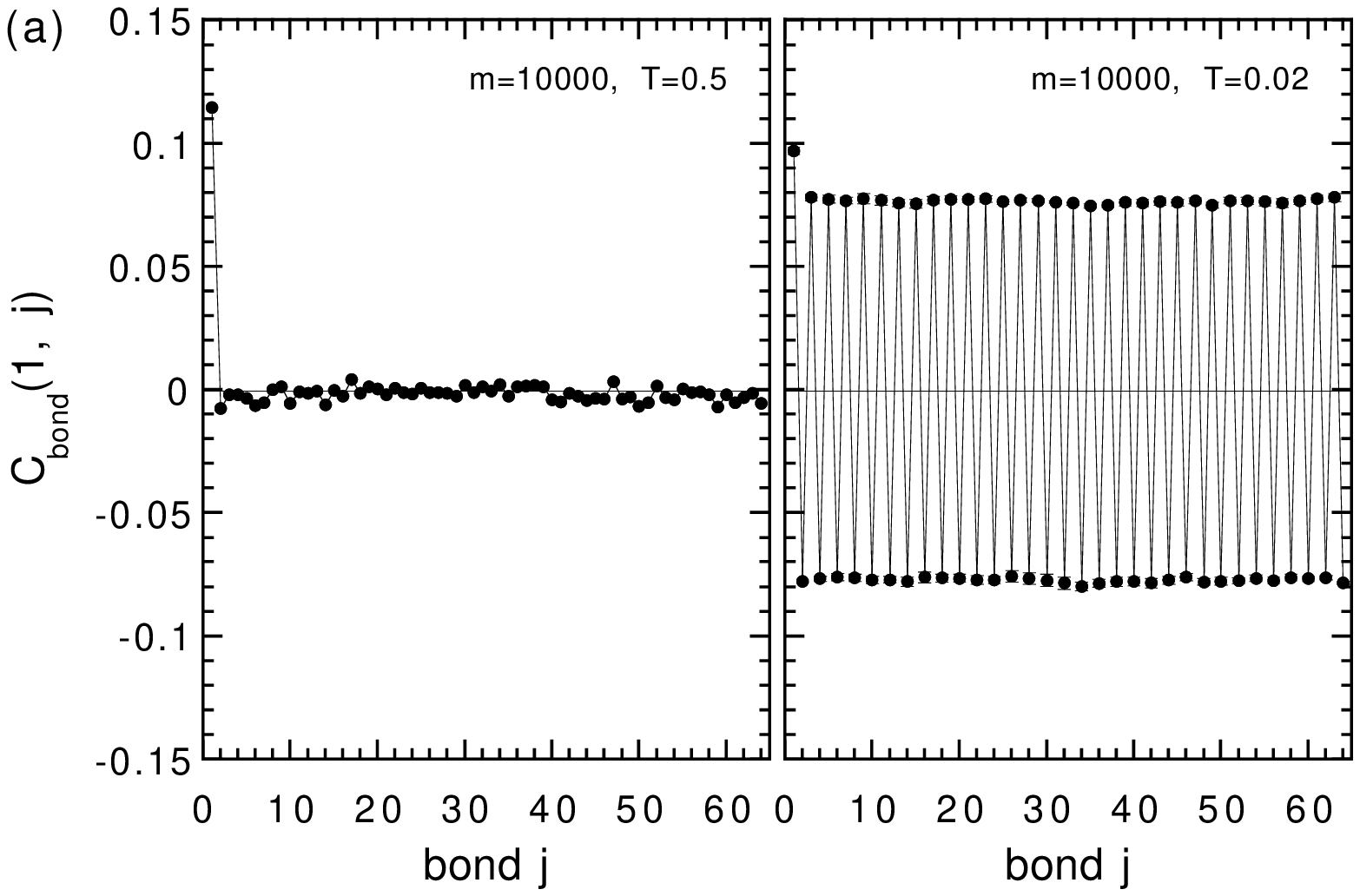}
\includegraphics[width=80mm]{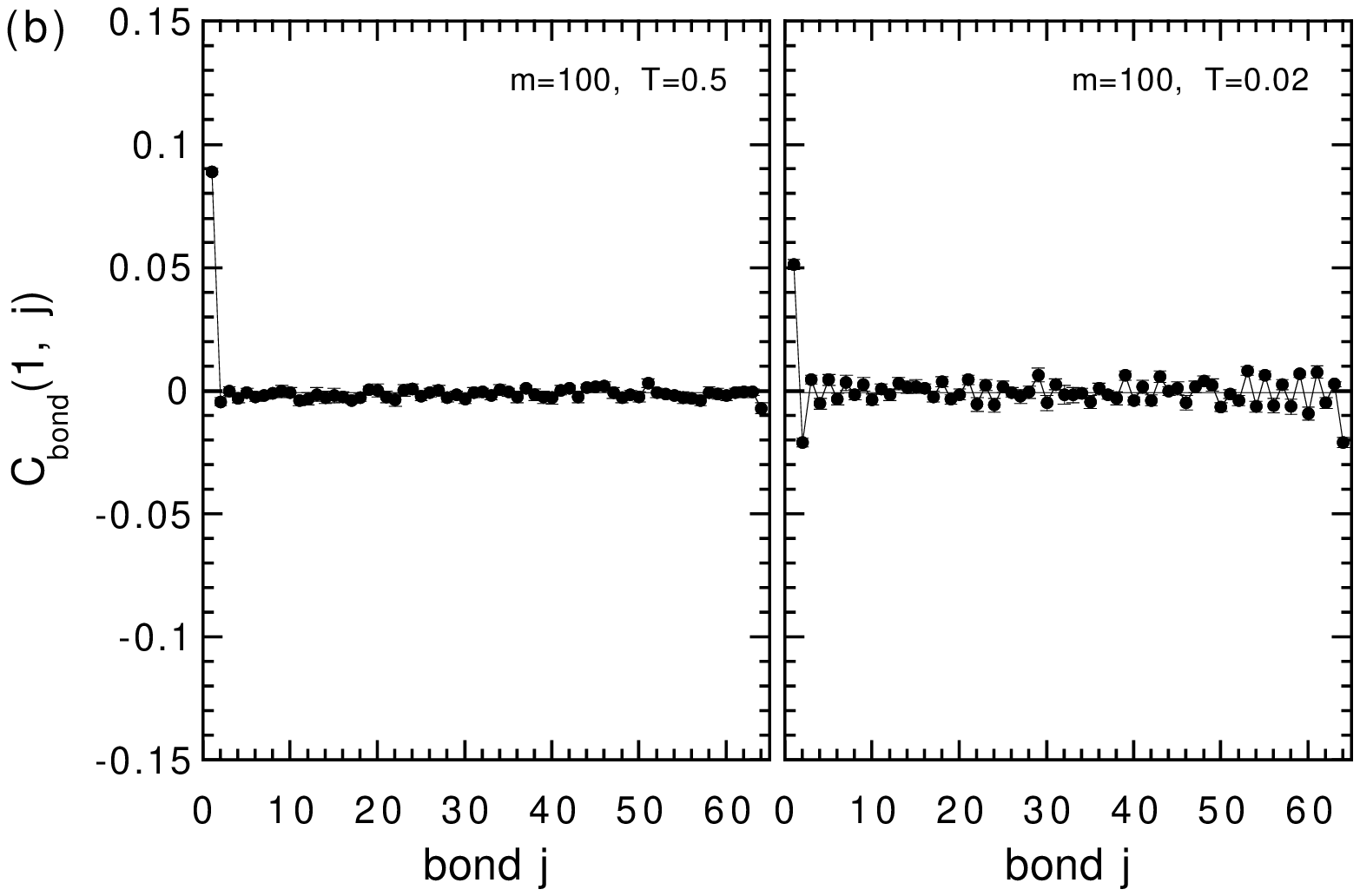}
\includegraphics[width=80mm]{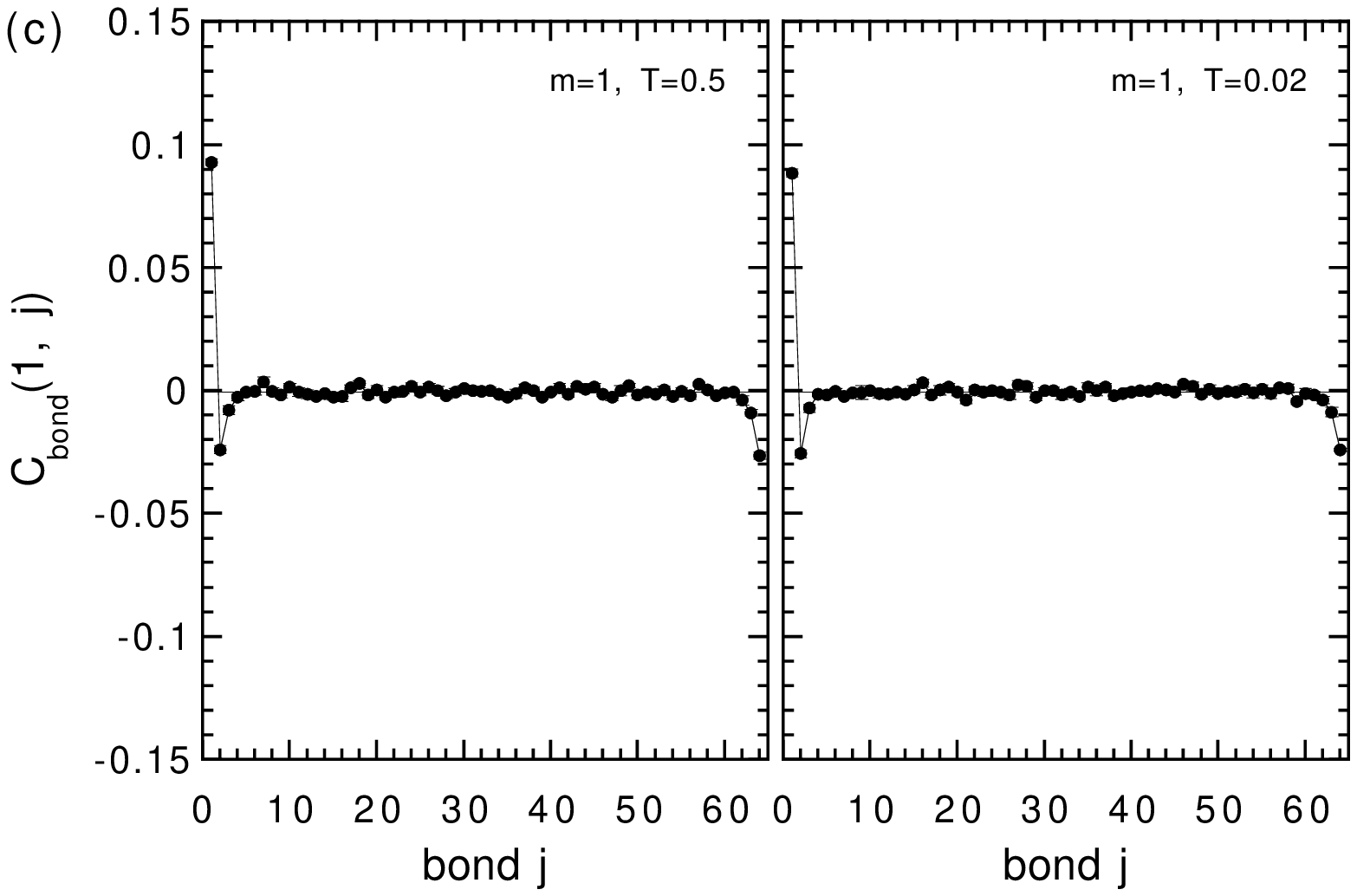}
\end{center}
\caption{%
The bond correlation function along the real-space axis for
(a) $m=10000$,
(b) $m=100$
and
(c) $m=1$.
The left figures are the data at $T=0.5$
for the system of $N=64$ and $M=12$.
The right figures are the data at $T=0.02$
for the system of $N=64$ and $M=384$.
}
\label{fig:bxcr}
\end{figure}

In Figs.~\ref{fig:bxcr}
we show the bond correlation function
from the left edge bond,
\begin{equation}
C_{\rm bond}(1,j) = \langle \Delta_1 \Delta_j \rangle,
\end{equation}
where $\langle \cdots \rangle$ denotes the thermal average.
For a heavy mass $m=10000$,
the bond correlation function depicted in Fig.~\ref{fig:bxcr}(a)
agrees with our previous results
without quantum fluctuation of the bond.~\cite{spqmc}
At a high temperature $T=0.5$,
each bond fluctuates with no correlation
and the bond takes a uniform configuration on the thermal average
due to thermal fluctuation of the bond.
At a very low temperature $T=0.02$,
the bond alternation occurs
with strong correlation extending over the chain.
When the mass decreases,
the effect of quantum lattice fluctuation becomes remarkable.
As we show in Figs.~\ref{fig:bxcr}(b) and \ref{fig:bxcr}(c),
the alternating structure at low temperature disappears.
We find no alternating pattern for a light mass $m=1$,
where the bond takes a uniform configuration
because the bond alternation is disarranged
by quantum fluctuation of the bond.

%
%
\begin{figure}[t]
\begin{center}
\includegraphics[width=80mm]{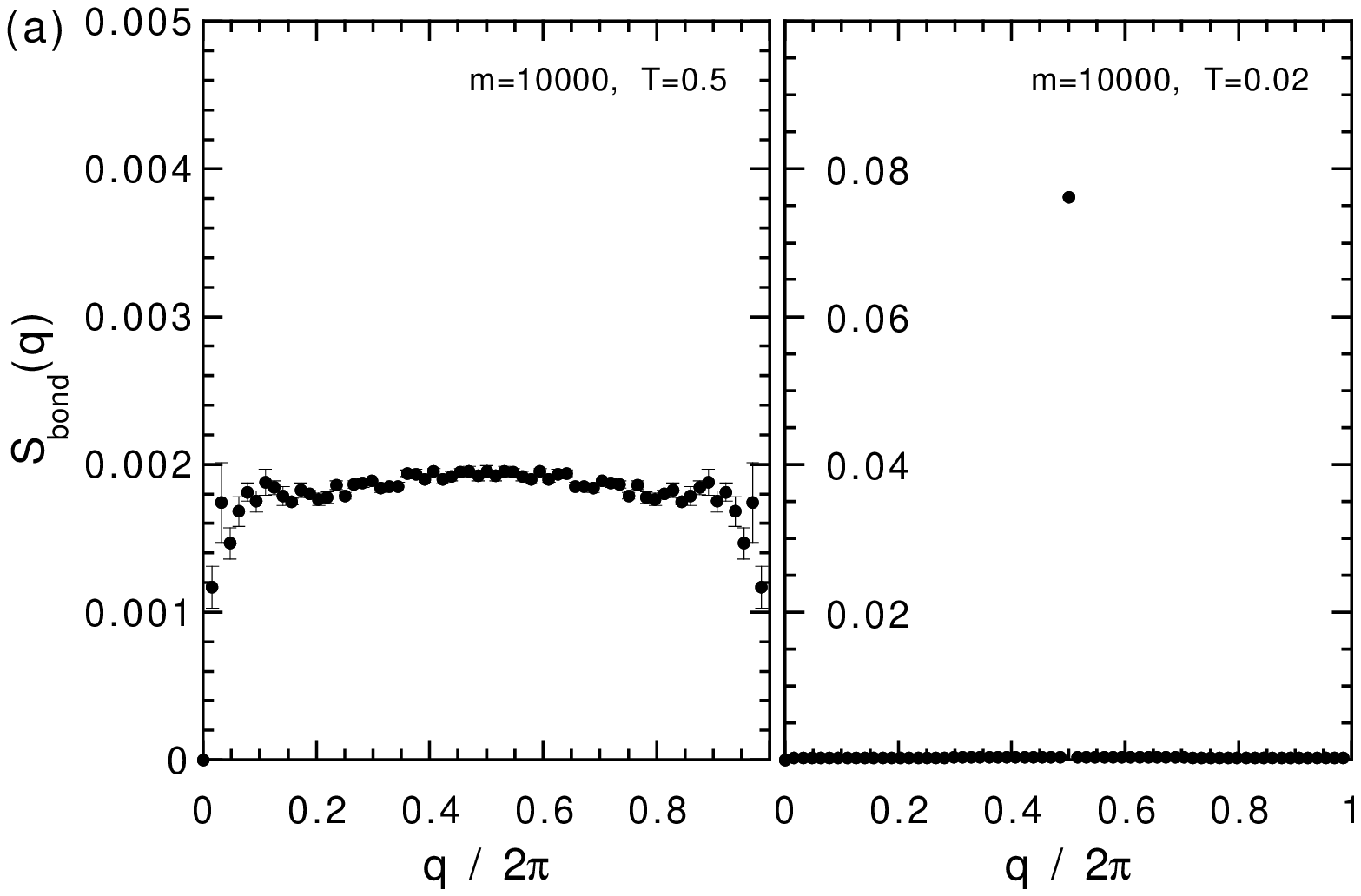}
\includegraphics[width=80mm]{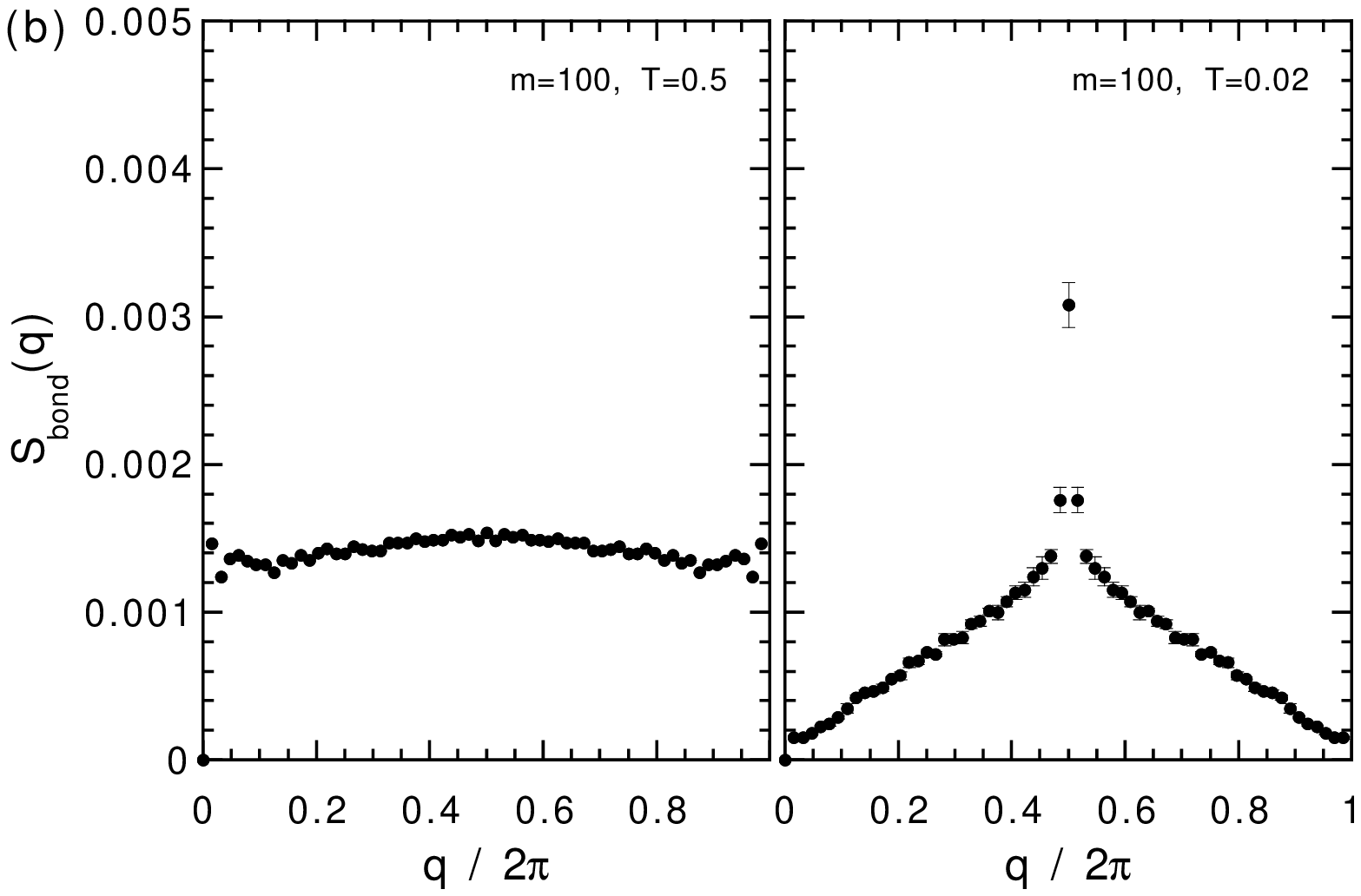}
\includegraphics[width=80mm]{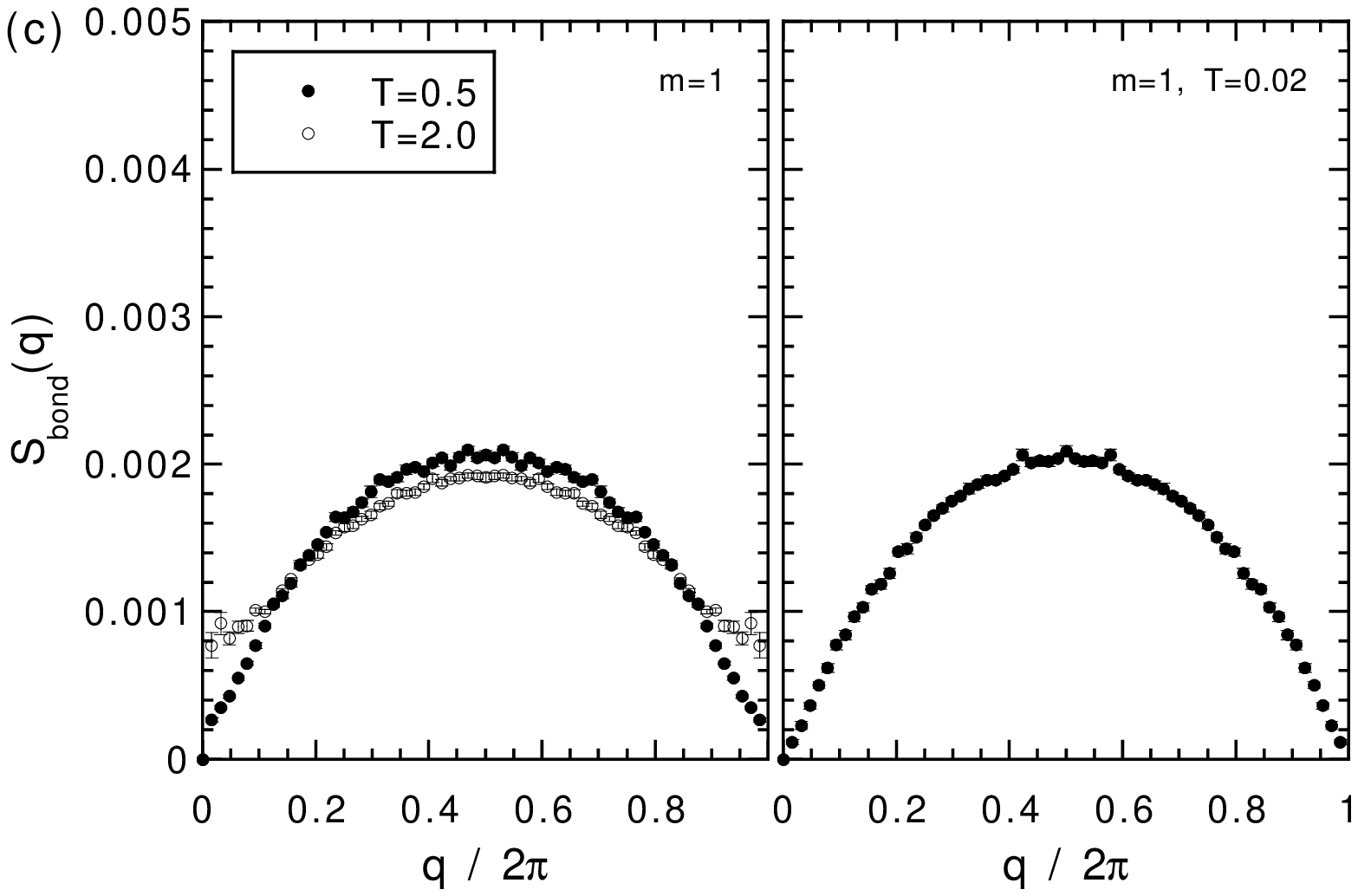}
\end{center}
\caption{%
The bond structure factor for
(a) $m=10000$,
(b) $m=100$
and
(c) $m=1$.
The left figures are the data at $T=0.5$
for the system of $N=64$ and $M=12$.
The right figures are the data at $T=0.02$
for the system of $N=64$ and $M=384$.
The open circle in (c) denotes the data at $T=2.0$
for the system of $N=64$ and $M=4$.
}
\label{fig:bq}
\end{figure}

Thus
the uniform bond configuration is realized due to
thermal fluctuation at high temperature
and due to quantum fluctuation at low temperature.
The shape of the bond correlation function is simply flat for both cases.
Therefore
the bond correlation function is insufficient
to characterize the difference of the nature of the bond fluctuation
in both cases.
For the purpose of
clarifying the distinction
between the classical bond fluctuation and the quantum one,
we investigate the bond structure factor,
\begin{equation}
S_{\rm bond}(q)
=
\langle |\Delta_q|^2 \rangle,
\end{equation}
where
\begin{equation}
\Delta_q = \frac{1}{N}\sum_{j=1}^N\Delta_j{\rm e}^{{\rm i}qj}.
\end{equation}
In Fig.~\ref{fig:bq}(a)
we show the bond structure factor for $m=10000$.
At $T=0.5$,
$S_{\rm bond}$ is flat
as a result of the mixture of various modes of the motion,
which means that each bond fluctuates independently.
At $T=0.02$,
the lattice forms the bond-alternate structure
and $S_{\rm bond}$ has a Bragg peak at $q=\pi$.
Except for $q=\pi$,
$S_{\rm bond}$ is very small
and the fluctuation is independent of the position of the bond.
These structures are consistent with the results
of the classical lattice system.

On the other hand,
when the mass decreases,
the bond structure factor shows
characteristic dependence on the temperature.
In Fig.~\ref{fig:bq}(b)
we show the bond structure factor for $m=100$.
$S_{\rm bond}$ remains flat at $T=0.5$.
At $T=0.02$,
comparing with the data for $m=10000$,
the amplitude of the Bragg peak becomes small
and the fluctuation of various modes grows in addition to $q=\pi$.
This structure factor with the mixture of various modes
corresponds to the correlation function
shown in the right of Fig.~\ref{fig:bxcr}(b),
where the alternate structure is hardly detected.

When the mass decreases furthermore, $m=1$,
$S_{\rm bond}$ has a sinusoidal shape
at both $T=0.5$ and $T=0.02$
as shown in Fig.~\ref{fig:bq}(c).
The sinusoidal structure factor is interpreted as follows.
For light mass,
the second term of eq.~(\ref{eq:hamsp})
is much larger than the first one.
Namely,
the lattice moves almost independently of the spin configuration.
Therefore,
in the following,
we investigate the lattice part of eq.~(\ref{eq:hamsp}),
\begin{equation}
H_{\rm phonon}=
\sum_{j=1}^{N}
\left[
\frac{1}{2m}p_{j}^2 + \frac{k}{2}(u_{j}-u_{j+1})^2
\right].
\label{eq:hamph}
\end{equation}
The operators of the lattice displacement $u_j$ and the momentum $p_j$
are expressed by
the creation operator $a_q^{\dag}$ and the annihilation operator $a_q$
of the phonon,
\begin{eqnarray}
u_{j}
&=&
\sum_{q} \sqrt{\frac{1}{2Nm\omega_{q}}}
(a_{-q}^{\dag}+a_{q}){\rm e}^{{\rm i}qj},
\\
p_{j}
&=&
\sum_{q} {\rm i}\sqrt{\frac{m\omega_{q}}{2N}}
(a_{-q}^{\dag}-a_{q}){\rm e}^{{\rm i}qj},
\end{eqnarray}
where
\begin{equation}
\omega_{q}^2 = 4\omega^{2}\sin^{2}\frac{q}{2}.
\end{equation}
Then eq.~(\ref{eq:hamph}) is transformed into
\begin{equation}
H_{\rm phonon} = \sum_{q} \omega_{q}\left( a^{\dag}_{q}a_{q}+\frac{1}{2}\right).
\end{equation}
The Fourier component of the bond distortion is expressed by
\begin{equation}
\Delta_q =
2{\rm i}\alpha\sqrt{\frac{1}{2Nm\omega_{q}}}
(a_{q}^{\dag}+a_{-q}){\rm e}^{-{\rm i}q/2}\sin\frac{q}{2}.
\end{equation}
At $T=0$,
all the nomal modes are in the ground state,
i.e., $a_{q}|0\rangle=0$,
and $S_{\rm bond}$ is given by
\begin{eqnarray}
S_{\rm bond}(q)
&=&
\langle |\Delta_q|^2 \rangle
\nonumber\\
&=&
\frac{2\alpha^2}{Nm\omega_q}\sin^2\frac{q}{2}
\langle 0|
(a_{-q}^{\dag}+a_{q})
(a_{q}^{\dag}+a_{-q})
|0 \rangle
\nonumber\\
&=&
\frac{\alpha^2}{Nm\omega}\sin\frac{q}{2}.
\label{eq:sqbond}
\end{eqnarray}
Namely,
the zero-point motion gives the sinusoidal structure factor.
This result agrees with the sinusoidal shape of our Monte Carlo results.
But we find that the amplitude in Fig.~\ref{fig:bq}(c) ($\cong 0.0021$) is smaller than
that of eq.~(\ref{eq:sqbond}) ($\cong 0.0156$).
This discrepancy is caused by the cutoff for the bond distortion.
The bond distortion is suppressed due to the cutoff
and the structure factor becomes small.
In Monte Carlo simulations of the lattice part without the spin part,
it is confirmed that
the amplitude agrees with that of eq.~(\ref{eq:sqbond})
in the case where the cutoff is large enough,
and
the amplitude agrees with that of Fig.~\ref{fig:bq}(c)
in the case where the cutoff is the same as
that we provide in the spin-Peierls model.
Thus
we conclude that
the lattice fluctuation for light mass is characterized by
the zero-point quantum fluctuation of the lattice.
It should be mentioned that
when the temperature is larger than $\omega=\sqrt{k/m}$,
$S_{\rm bond}$ is influenced by the excited state of the phonon.
In the present case,
we actually find that
$S_{\rm bond}$ is influenced by excited states at $T=2.0$ ($>\omega=1.0$)
and changes from the sinusoidal shape to a flat shape
as shown by the open circle in the left of Fig.~\ref{fig:bq}(c).

\section{Quantum Lattice Fluctuation}
\label{sec:quantum lattice fluctuation}

\begin{figure}[t]
\begin{center}
\includegraphics[width=80mm]{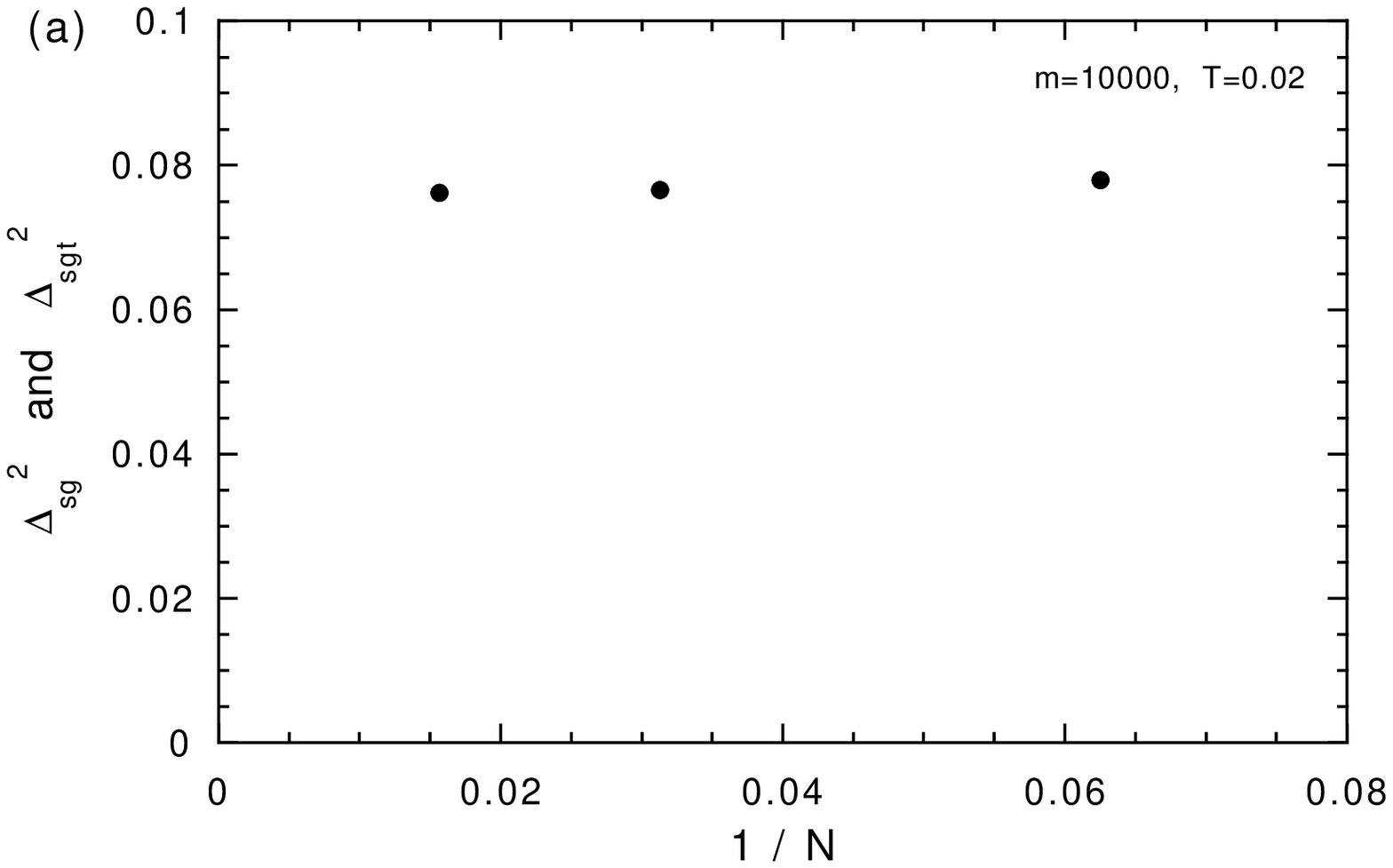}
\includegraphics[width=80mm]{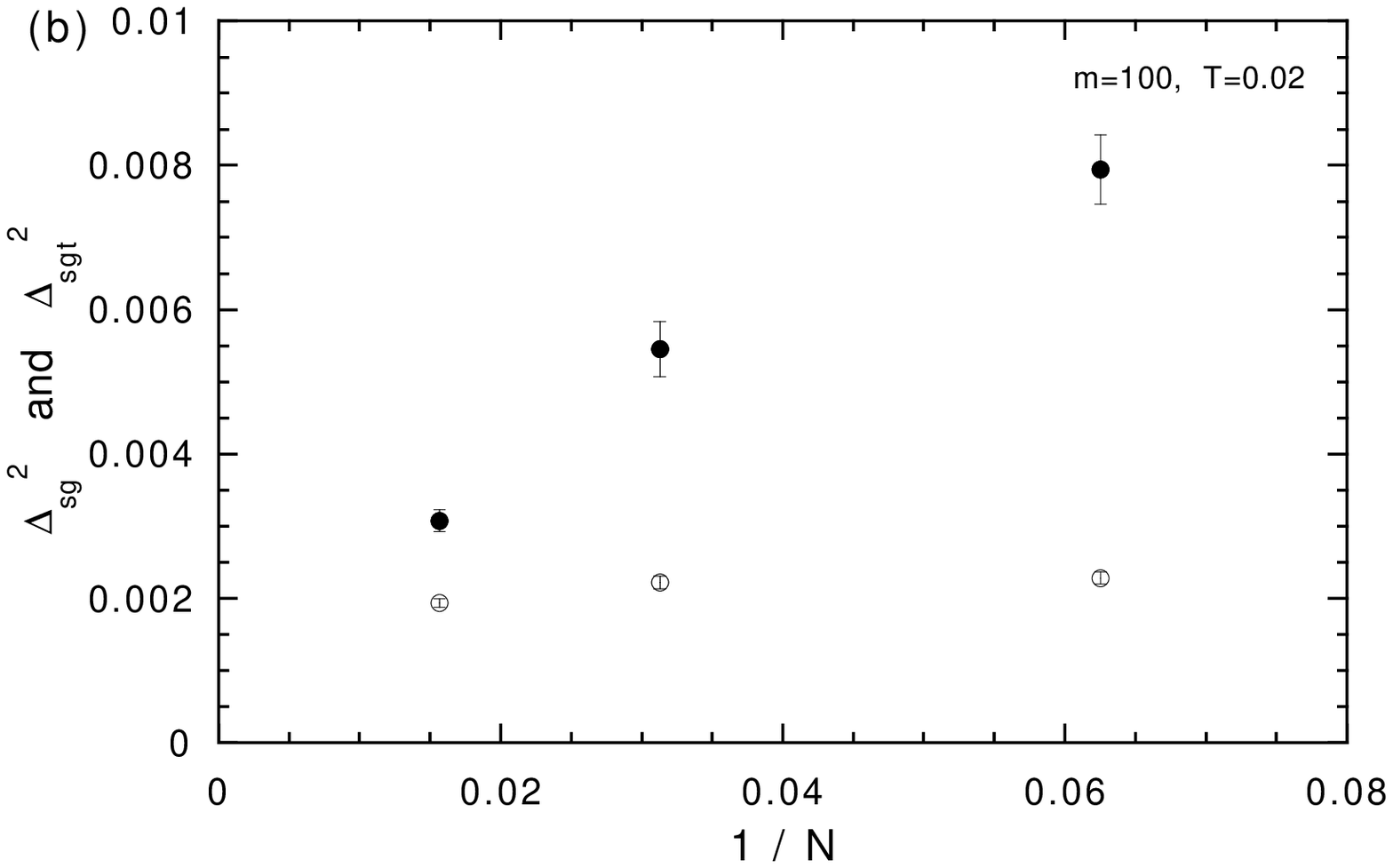}
\includegraphics[width=80mm]{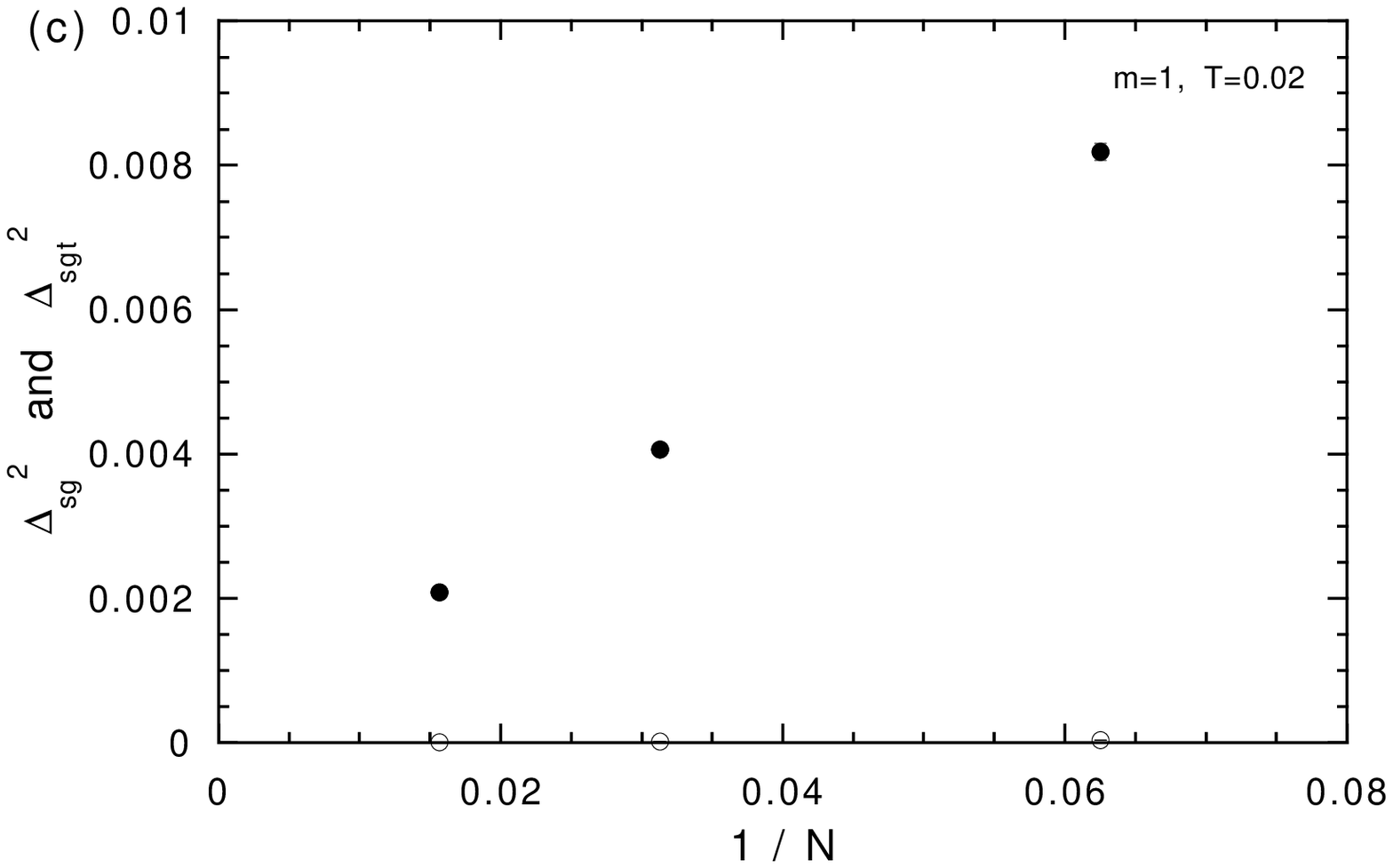}
\end{center}
\caption{%
The size dependence of the order parameters for the bond alternation
$\Delta_{\rm sg}^2$ (solid circle)
and $\Delta_{\rm sgt}^2$ (open circle) for
(a) $m=10000$,
(b) $m=100$
and
(c) $m=1$
at $T=0.02$. $M=384$.
}
\label{fig:bsg2}
\end{figure}

In the thermal average,
the lattice takes either the uniform configuration for light mass
or the dimerized configuration for heavy mass
due to the spin-phonon coupling.
On the other hand,
in a snapshot in the Monte Carlo simulation,
the world line of the lattice
shows various configurations due to quantum lattice fluctuation.
For heavy mass,
the motion of the lattice is regarded to be adiabatic
and the world-line configuration of the lattice is straight
along the imaginary-time axis.
The straight world line parallels with each other
and forms the dimerized configuration.
As the mass decreases,
quantum lattice fluctuation becomes relevant and
the world line of the lattice begins to fluctuate
along the imaginary-time axis.
The deviation of the world line
may suppress the bond-alternating order.
Here we introduce the order parameter for the bond-alternating order,
\begin{equation}
\Delta_{\rm sg}^2
=
\left\langle
\left(
\frac{1}{N}
\sum_{i=1}^{N}(-1)^i\Delta_{i}
\right)^2
\right\rangle.
\end{equation}
In addition,
we explore how the bond alternation is disarranged
due to quantum lattice fluctuation
by investigating the following quantity,
\begin{equation}
\Delta_{\rm sgt}^2
=
\left\langle
\left(
\frac{1}{\beta}\int_{0}^{\beta}{\rm d}\tau
\frac{1}{N}
\sum_{i=1}^{N}(-1)^i
\, {\rm e}^{H\tau}
\Delta_{i}
\, {\rm e}^{-H\tau}
\right)^2
\right\rangle,
\end{equation}
where ${\rm e}^{H\tau}\Delta_{i}\,{\rm e}^{-H\tau}$ represents
the value of $\Delta_{i}$ at the imaginary time $\tau$.
In Figs.~\ref{fig:bsg2}
we show the size dependence
of $\Delta_{\rm sg}^2$ and $\Delta_{\rm sgt}^2$ at $T=0.02$.
For $m=10000$,
$\Delta_{\rm sg}^2$ agrees with $\Delta_{\rm sgt}^2$,
which indicates that
the bond-alternating order develops
in both the real-space axis and the imaginary-time axis.
In fact,
there the world line is straight along the imaginary-time axis.
When the mass decreases,
the absolute values of
the bond-alternating correlations decrease
as shown in Figs.~\ref{fig:bsg2}(b) and \ref{fig:bsg2}(c).
In particular,
$\Delta_{\rm sgt}^2$ is much reduced comparing with $\Delta_{\rm sg}^2$
for finite size systems.
This is due to the deviation of the world-line configuration
along the imaginary-time axis.
For $m=100$, however,
we find that
$\Delta_{\rm sg}^2$ and $\Delta_{\rm sgt}^2$
converge to almost the same finite value
in the thermodynamic limit.
The amount of the bond alternation for $m=100$
is smaller than that for $m=10000$.
For $m=1$,
the bond configuration is uniform on the thermal average
in the thermodynamic limit.
There
$\Delta_{\rm sg}^2$ shows $1/N$ dependence
which indicates a short range order of the bond alternation,
while
$\Delta_{\rm sgt}^2$ is almost zero
because the correlation length along the imaginary-time axis
is much shorter than $\beta$
as we will see in
\S\ref{sec:quantum narrowing effect}.

%
%

\section{Mass Dependence of Magnetic Properties}
\label{sec:mass dependence of magnetic properties}

\begin{figure}[t]
\begin{center}
\includegraphics[width=80mm]{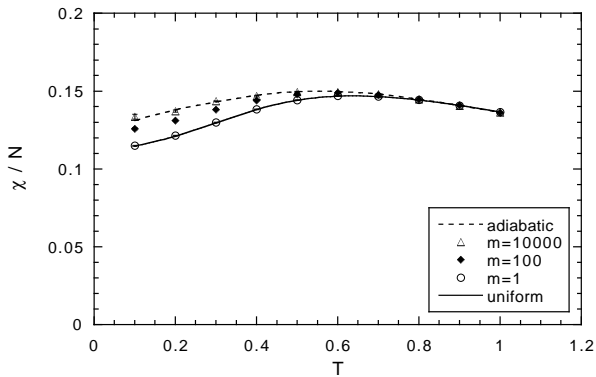}
\end{center}
\caption{%
The temperature dependence of the magnetic susceptibility.
The data are calculated for the system of $N=64$
and extrapolated with respect to the Trotter number.
The dotted line denotes the data for the adiabatic lattice system of $N=64$
and
the solid line denotes the data for the uniform lattice system of $N=64$.
}
\label{fig:xmg}
\end{figure}

The spin-phonon coupling causes deviation of magnetic properties
from those of the uniform lattice system.
First
we consider the effect of thermal fluctuation of the lattice
at high temperature where the spin-Peierls correlation is not relevant.
In Fig.~\ref{fig:xmg}
we show the temperature dependence of the magnetic susceptibility
for a high temperature region.
We have calculated for the adiabatic lattice system.~\cite{spqmc}
For comparison,
the data
for the adiabatic lattice system (dotted line)
and the uniform lattice system (solid line)
are shown in the figure.
For $m=10000$,
the magnetic susceptibility agrees with that of the adiabatic lattice system.
As has been pointed out in numerical works,~\cite{ph-qmc1,ph-qmc3,spqmc}
the magnetic susceptibility deviates from that of the uniform lattice system
even at high temperatures
due to thermal fluctuation of the lattice.
As the mass decreases,
the magnetic susceptibility approaches
that of the uniform lattice system monotonously.
For $m=1$,
the magnetic susceptibility agrees with
that of the uniform lattice system well,
even though the lattice strongly fluctuates.
This fact has been seen
in previous works
for models with the second quantization of the lattice.~\cite{ph-qmc1,ph-qmc3}

%
%
%
%
\begin{figure}[t]
\begin{center}
\includegraphics[width=80mm]{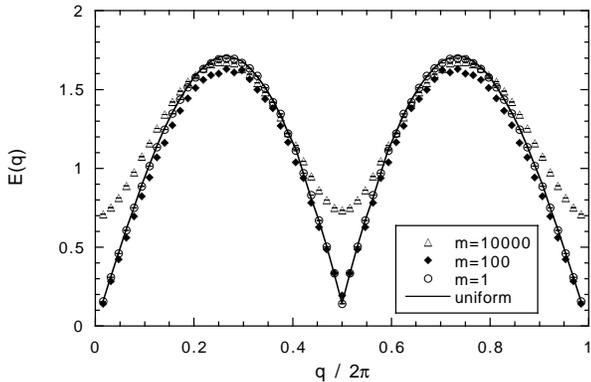}
\end{center}
\caption{%
The magnetic elementary excitation spectrum in the ground state.
The data are calculated at $T=0.02$ for the system of $N=64$ and $M=384$.
The solid line denotes the data for the uniform lattice system of $N=64$.
}
\label{fig:excitation}
\end{figure}

Now
we consider the effect of quantum fluctuation of the lattice
at low temperature.
In order to grasp the magnetic behavior from a microscopic point of view,
we investigate the magnetic excitation spectrum.
It can be extracted from
the imaginary-time correlation function
of the spin,~\cite{spectrum-1,spectrum-2}
\begin{equation}
S_{\rm spin}(q,\tau)
=
\langle
{\rm e}^{H\tau}S_q^z
{\rm e}^{-H\tau}S_{-q}^z
\rangle,
\label{eq:sqtau}
\end{equation}
where
\begin{equation}
S_q^z
=
\frac{1}{N}\sum_{j=1}^N S_j^z{\rm e}^{{\rm i}qj}.
\end{equation}
The dispersion relation $E(q)$ of
the elementary excitation is obtained by
\begin{equation}
E(q)=-\frac{\partial}{\partial\tau}\ln [S_{\rm spin}(q,\tau)]
\end{equation}
for large $\tau$.
In Fig.~\ref{fig:excitation}
we show the magnetic elementary excitation spectrum at $T=0.02$
for several values of the mass.
The solid line denotes the data
for the uniform lattice system of $N=64$,
where we find a small energy gap
due to the finite size effect.
For $m=10000$,
the lattice motion is adiabatic
and the bond alternation occurs at low temperature.
Accordingly,
we find an energy gap at $q=\pi$ clearly.
On the other hand,
we find that
the energy gap becomes small drastically
as the mass decreases.
For $m=1$,
$E(q)$ coincides with that of the uniform lattice system.
Thus we find that
not only the magnetic susceptibility
but also the microscopic magnetic characteristic
agrees with those of the uniform lattice system.
Therefore
we conclude that this agreement is intrinsic.
In the next section,
we study the physical background of this agreement.

\section{Quantum Narrowing Effect}
\label{sec:quantum narrowing effect}

%
%
\begin{figure}[t]
\begin{center}
\includegraphics[width=80mm]{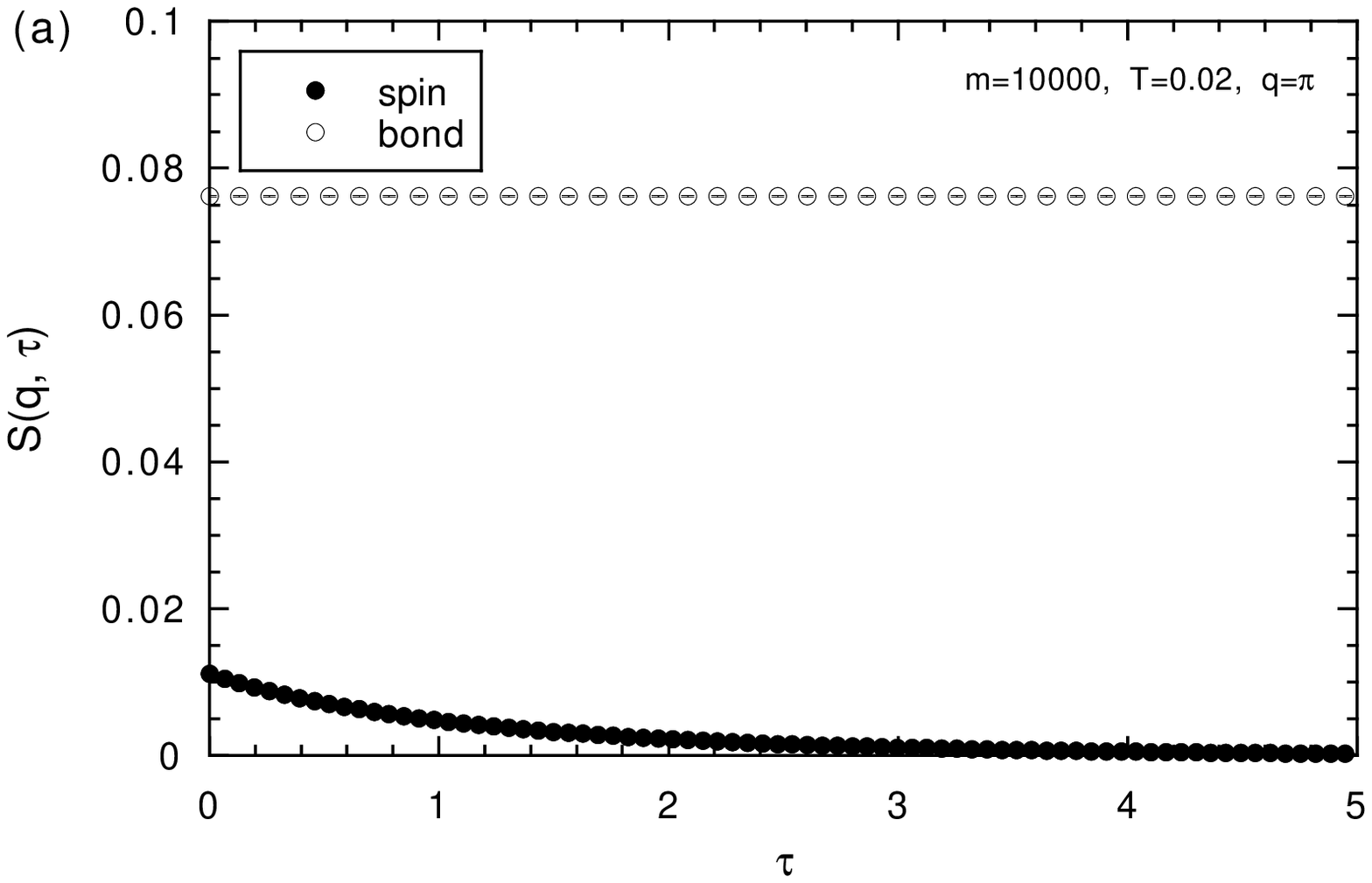}
\includegraphics[width=80mm]{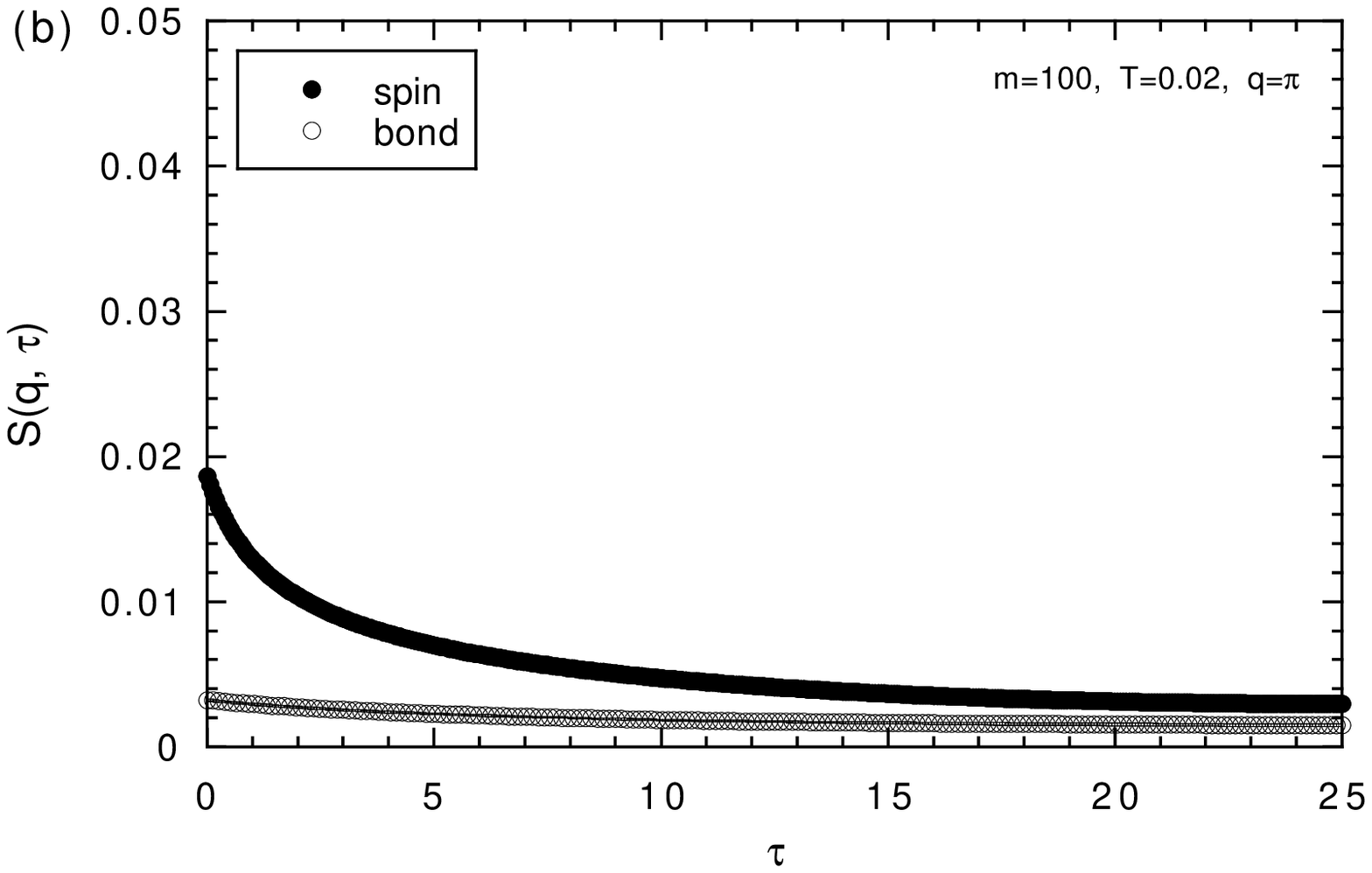}
\includegraphics[width=80mm]{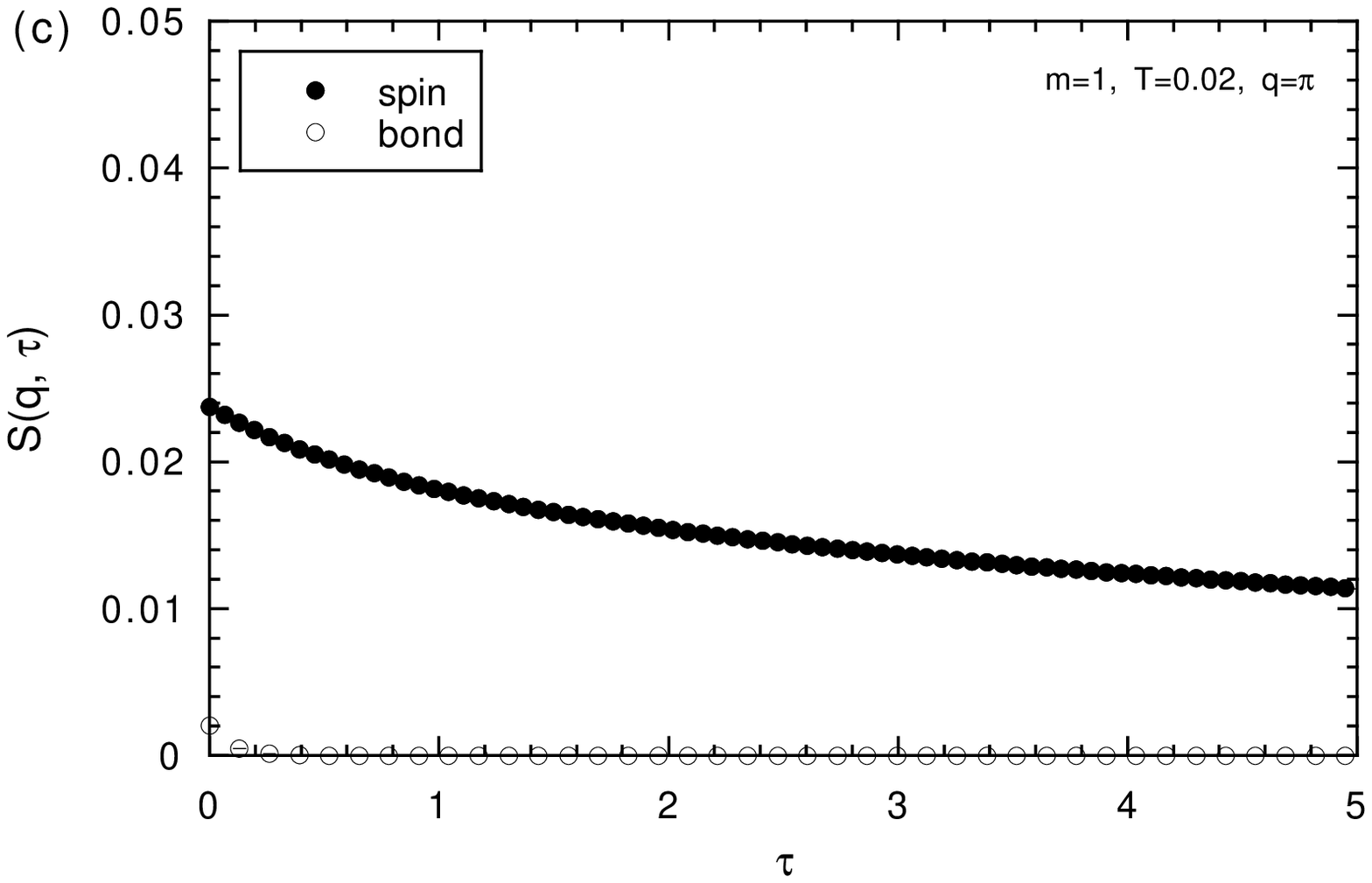}
\end{center}
\caption{%
The imaginary-time correlation functions of the spin and the bond
for $q=\pi$ at $T=0.02$.
(a) $m=10000$,
(b) $m=100$
and
(c) $m=1$.
The solid circle and the open circle denote
the data in terms of the spin and the bond, respectively.
The data are calculated for the system of $N=64$ and $M=384$.
}
\label{fig:sbtcr}
\end{figure}

In this section
we consider the reason why
the expectation value of magnetic properties of the light mass system
coincides with that of the uniform lattice system.
As we mentioned above,
the motion of the lattice is almost independent of the spin configuration
because the mass is light.
On the other hand,
the motion of the spin may be affected by the bond fluctuation.
Generally speaking, however,
the effect of fast fluctuation is to be neglected,
which is known as the motional narrowing effect.
Namely,
when the fluctuation is much fast
compared to the time scale which we would like to observe,
the fluctuation is leveled off
and does not affect other degrees of freedom.
In the present case,
what comes into question is the degree of quantum fluctuations
of the spin and the lattice.
The correlation along the imaginary-time axis
gives information on the time scale of quantum fluctuation.
We investigate
$S_{\rm spin}(q,\tau)$ defined in eq.~(\ref{eq:sqtau})
and the imaginary-time correlation function of the bond,
\begin{equation}
S_{\rm bond}(q,\tau)
=
\langle
{\rm e}^{H\tau}\Delta_q
{\rm e}^{-H\tau}\Delta_{-q}
\rangle.
\label{eq:bqtau}
\end{equation}
In Figs.~\ref{fig:sbtcr}
we compare both imaginary-time correlation functions for $q=\pi$ at $T=0.02$.
For $m=10000$,
$S_{\rm spin}(q,\tau)$ decays along the imaginary-time axis quickly
and $S_{\rm bond}(q,\tau)$ keeps a constant value.
The spin fluctuation is much faster than the bond fluctuation.
Namely,
the time scale of the spin fluctuation $\tau_{\rm spin}$ is much smaller than
that of the bond fluctuation $\tau_{\rm bond}$
($\tau_{\rm spin} \ll \tau_{\rm bond}$).
In this case,
the lattice vibration can be regarded to be adiabatic.
On the other hand,
as the mass decreases,
the inequality relation between the two time scales is reversed.
For $m=1$,
$S_{\rm bond}(q,\tau)$ decays quite rapidly.
This value of $\tau_{\rm bond}$ is much shorter than $\beta=(0.02)^{-1}=50$
as we expected by the observation of $\Delta_{\rm sgt}^2$
in \S\ref{sec:quantum lattice fluctuation}.
Here
the relaxation of $S_{\rm bond}(q,\tau)$
is much faster than that of $S_{\rm spin}(q,\tau)$,
i.e., $\tau_{\rm spin} \gg \tau_{\rm bond}$.
This means that
the lattice fluctuates much faster
than the time scale of the motion of the spin.
Thus we conclude that
a sort of narrowing effect happens
and the lattice is effectively fixed to be uniform.
We call this phenomenon
{\lq\lq}quantum narrowing effect{\rq\rq}.

\section{Summary and Discussion}
\label{summary and discussion}

In this paper
we investigated
a one-dimensional $S=1/2$ antiferromagnetic Heisenberg model
coupled to quantum lattice vibration
using a quantum Monte Carlo method.
We studied
the mass dependence of various physical quantities.
For heavy mass,
quantum lattice vibration is regarded as the adiabatic motion
and the system dimerizes at low temperature.
When the mass decreases,
the bond-alternate correlation is reduced
due to the deviation of the world-line configuration of the lattice.
For light mass,
the spin-Peierls correlation does not grow even at low temperature,
and
the ground-state lattice fluctuation is equivalent to
the zero-point vibration of the phonon.
The lattice fluctuation does not affect
the magnetic behavior.
We found the separation of time scales
of quantum fluctuations of spin ($\tau_{\rm spin}$)
and lattice ($\tau_{\rm bond}$) degrees of freedom.
The time scales change with the mass.
The motion of the lattice can be regarded to be adiabatic
for heavy mass,
where $\tau_{\rm spin}\ll\tau_{\rm bond}$.
On the other hand,
the motion of the lattice is smeared out due to the narrowing effect
for light mass,
where $\tau_{\rm spin}\gg\tau_{\rm bond}$.

For intermediate mass,
$m=100$ [Fig.~\ref{fig:sbtcr}(b)],
imaginary-time correlation functions of the spin and the bond
decay with almost the same decay time.
In such cases
we expect that
the spin fluctuation and the bond fluctuation resonate with each other.
The resonance of quantum fluctuations
would cause a multiple quantum state of spin and lattice degrees of freedom,
which is an interesting problem in the future.

\section*{Acknowledgements}

The present study is partially supported by
a Grant-in-Aid from the Ministry of Education, Culture, Sports,
Science and Technology of Japan.
The authors are also grateful for the use of the Supercomputer Center
at the Institute for Solid State Physics, University of Tokyo.
Part of the calculation has been done
in the computer facility
at Advanced Science Research Center,
Japan Atomic Energy Research Institute.


\end{document}